\documentclass[twocolumn,english,aps,superscriptaddress]{revtex4}

\usepackage{amssymb}
\usepackage{amsmath}
\usepackage{colordvi}
\usepackage[colorlinks]{hyperref}
\usepackage{amsthm}
\usepackage{subeqnarray}
\usepackage{cases}
\usepackage{enumerate}
\usepackage{graphicx}
\usepackage{epsfig}
\usepackage{epstopdf}
\usepackage{subfigure}
\usepackage{color}
\usepackage{url}
\usepackage{verbatim}
\usepackage{mathrsfs}

\usepackage{color}

\begin{document}

\title{Long-range interactions and symmetry-breaking in quantum gases through optical feedback}
\author{Yong-Chang Zhang}
\email{yczhang@phys.au.dk}
\author{Valentin Walther}
\author{Thomas Pohl}
\affiliation{Department of Physics and Astronomy, Aarhus University, Ny Munkegade 120, 8000 Aarhus C, Denmark}

\begin{abstract}
We consider a quasi two-dimensional atomic Bose Einstein condensate interacting with a near-resonant laser field that is back-reflected onto the condensate by a planar mirror. We show that this single-mirror optical feedback leads to an unusual type of effective interaction between the ultracold atoms giving rise to a rich spectrum of ground states. In particular, we find that it can cause the spontaneous contraction of the quasi two-dimensional condensate to form a self-bound one-dimensional chain of mesoscopic quantum droplets, and demonstrate  that the observation of this exotic effect is within reach of current experiments.
\end{abstract}

\maketitle

Ultracold atomic gases have emerged as a unique platform for exploring the rich physics of quantum many-body systems in and out of equilibrium \cite{Bloch1,Gross,Polkovnikov,Nandkishore,Langen}. These new capabilities are based on the exquisite control of cold atoms by external fields and in particular the ability to control and tune atomic interactions. Most prominently, the use of Feshbach resonances \cite{Chin}, induced by magnetic or optical fields, makes it possible to vary the strength of zero-range collisional interactions between cold atoms, which is essential for their use in quantum simulations. The generation and control of finite-range interactions is significantly expanding the scope of such capabilities and therefore has been a frontier of cold atom research in recent years. Among the most promising systems are cold ion crystals \cite{roos}, cold molecules \cite{Carr}, Rydberg atoms \cite{saffman_rev} and dipolar quantum gases \cite{santos2003}, which have lead to a number of recent breakthroughs, from the  simulation of quantum spin models \cite{schauss2012,schauss2015,Zeiher,Labuhn,zeiher2017,Bernien} to the discovery of exotic quantum-fluid phenomena \cite{pfau1,pfau2,ferlaino2016}.
Alternatively, effective interactions can be induced by another quantum system coupled to the cold atoms. Examples include polaron interactions in degenerate gases \cite{Devreese,Camacho1,Camacho2} and photon-mediated interactions induced by interfacing atoms with nano-scale photonic structures \cite{Petersen, Douglas} or optical cavities \cite{Baumann,Mottl,Habibian,Schmidt,Jager,Landig,Flottat,Leonard,Vaidya}.
Such and similar atom-photon interfaces offer unprecedented control capabilities for the emerging interactions \cite{Jager,Vaidya,Douglas}, and provide a unique platform for observing new phases of quantum matter \cite{Mottl,Habibian,Schmidt,Landig,Leonard,Piazza,Mivehvar} and fascinating collective opto-mechanical phenomena \cite{Ostermann,Schmittberger,Labeyire}.

\begin{figure}[b!]
\centering
  \includegraphics[width=0.99\columnwidth]{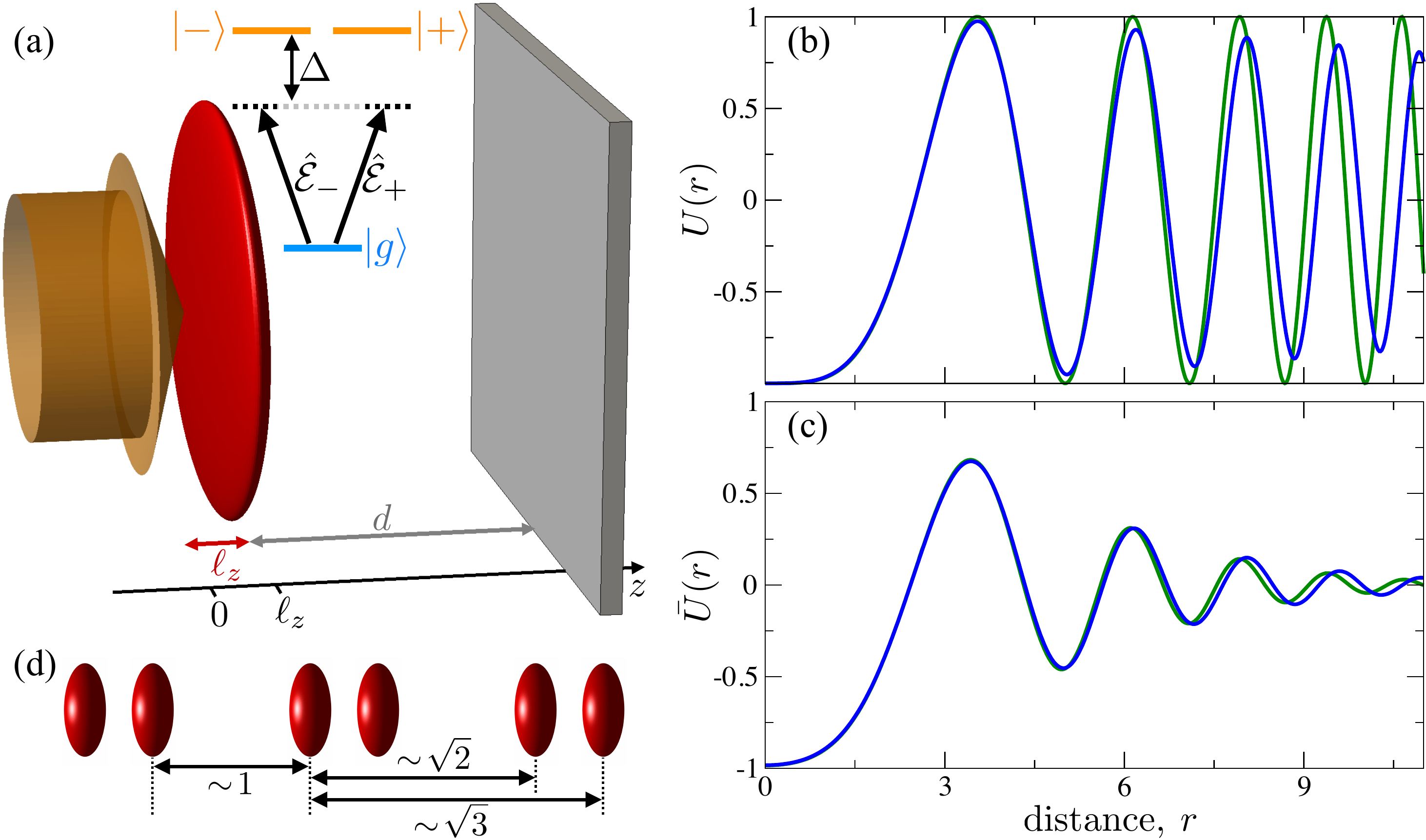}
\caption{(color online) (a) Schematic illustration of the considered setup in which a quasi-2D BEC is illuminated by a laser light that is back-reflected by a mirror and thereby generates optical feedback in the degenerate atom cloud. The atoms are driven by the incoming ($\hat{\mathcal{E}}_+$) and back-reflected ($\hat{\mathcal{E}}_-$) light field according to the depicted level scheme. The resulting multiple light scattering generates an effective atomic interaction which is shown in panel (b). The blue and green line depict the potential for $kd=20$ and in the limit $kd\rightarrow \infty$, respectively, where $d$ is the distance between the BEC and the mirror and $k=2\pi/\lambda$ is the wavenumber of the optical field. Panel (c) shows the potential energy resulting from the average interaction between two Gaussian clouds separated by a distance $r$, both with a RMS-radius of $\sigma=0.5$. The peculiar form of the interaction can stabilize a 1D lattice of quantum droplets that is illustrated in panel (d), along with the typical lattice spacings. All graphs use scaled dimensionless units described in the text.}
\label{fig1}
\end{figure}

Here, we consider one of the simplest possible setups, namely a light field coupled to an atomic gas via a mirror [see Fig.\ref{fig1}(a)]. We show that such a single-mirror optical feedback \cite{Ackemann,Labeyire,Tesio,Tesio1,Robb,Camara,Firth3,Moriya} induces an unusual type of atomic interaction that gives rise to equally exotic collective behavior of the optically driven quantum gas. In particular, we find that it can cause the spontaneous contraction of a quasi two-dimensional Bose Einstein condensate (BEC) into a self-bound one-dimensional chain of mesoscopic quantum droplets. Despite the simplicity of the setup, the system's ground state phase diagram indicates rich behavior, from quantum-droplet solutions and the aforementioned droplet-chains to extended density-wave states. The found structures exhibit an intrinsic emerging length scale and, in contrast to purely repulsive systems, such as dipolar condensates \cite{pfau1} or similar Rydberg-atom systems \cite{Hsueh,Cinti1,Cinti2}, are self-stabilized and do not rely on external in-plane confinement. Numerical simulations of the condensate dynamics under realistic conditions and including scattering-induced atom losses show that such states should be observable with present experimental capabilities.

As illustrated in Fig.\ref{fig1}{(a), we consider a quasi-2D BEC placed in front of a mirror \cite{Camara,Robb} at a distance $d$. Incident laser light propagates through the condensate, is back-reflected from the mirror and then traverses the BEC a second time. It is the optical feedback associated with the second encounter that can generate a nonlinear effect for the atoms. In classical terms, the light beam's effect can be intuitively understood as follows. As the light field propagates through the atomic cloud, it imparts an energy shift onto the atoms and at the same time acquires a phase shift which depends on the optical depth and therefore the local density of the atoms. Diffraction of the beam on its way to and from the mirror partly converts the beam's phase modulation into an amplitude modulation. The back-reflected modulated intensity, therefore, generates a light shift that depends on the atomic density. This opto-mechanical nonlinearity was found to trigger a dynamical instability and pattern formation for sufficiently high laser intensities \cite{Firth3,Tesio,Tesio1,Robb}, as observed in experiments with atomic vapor \cite{Labeyire,Camara}.

In order to formulate a quantum theory in the limit of small light intensities, we describe the forward and backward propagating fields by the slowly-varying electric field operators \cite{QuantumOpt} $\hat{\mathcal{E}}^\dagger_+$ and $\hat{\mathcal{E}}^\dagger_-$, respectively. These two fields couple the electronic ground state ($|g\rangle$) of the condensate atoms to two excited states $|\pm\rangle$ without interference. Such a configuration can be realized by using circularly polarized light and placing a quarter-wave plate between the BEC and the mirror. Within the rotating wave approximation, the coupled atom-photon system can then be described by the following Heisenberg equations
\begin{subequations}
\small
\begin{align}
i\frac{\partial \hat{\Psi}_\pm}{\partial t} &=  - \left(\Delta  +  \frac{\hbar}{2m} ( \frac{\partial^2}{\partial z^2} + \nabla^2_\perp) \right) \hat{\Psi}_\pm  -  \frac{\mu}{2\hbar}\hat{\Psi}_g \hat{\mathcal{E}}_\pm  \label{eq1a} \\
i\frac{\partial \hat{\Psi}_g}{\partial t} &= -\frac{\hbar\nabla^2}{2m} \hat{\Psi}_g +g_{3D} \hat{\Psi}^\dagger_g\hat{\Psi}_g\hat{\Psi}_g -\frac{\mu}{2\hbar}\!\left( \hat{\mathcal{E}}^\dagger_+ \hat{\Psi}_+ + \hat{\mathcal{E}}^\dagger_- \hat{\Psi}_-  \right)\label{eq1b} \\
 i \frac{\partial \hat{\mathcal{E}}_\pm}{\partial z}&= \mp \frac{1}{2k} (\frac{\partial^2 }{\partial z^2} + \nabla^2_\perp) \hat{\mathcal{E}}_\pm\mp \frac{k \mu}{2 \varepsilon_0  } \left( \hat{\Psi}^\dagger_g \hat{\Psi}_\pm +\hat{\Psi}^\dagger_\pm \hat{\Psi}_g \right) \label{eq1c}
\end{align}
\label{eq1}%
\end{subequations}
for the two light fields, $\hat{\mathcal{E}}_\pm$, and the field operators, $\hat{\Psi}_{g}({\bf r},z)$ and $\hat{\Psi}_{\pm}({\bf r},z)$, of the condensate atoms in the three respective electronic states. Here, $z$ defines the light propagation axis and ${\bf r}=(x,y)$, such that $\nabla^2_\bot=\frac{\partial^2}{\partial x^2}+\frac{\partial^2}{\partial y^2}$. Moreover, $m$ denotes the atomic mass, $\varepsilon_0$ is the vacuum permittivity, $k$ is the wave number of the optical field, and the constant  $g_{3D}=\frac{4\pi \hbar a}{m}$ denotes the strength of the zero-range interaction between ground state atoms, due to collisions with a scattering length $a$. Additional interaction terms due to collisions involving excited-state atoms can be neglected, since their density is much smaller than that of the ground state atoms, as discussed below. For simplicity, but without loss of generality \footnote{The derivation can be straightforwardly generalized to different detunings and dipole matrix elements of the two transition, which will however lead to the same form of Eq.(\ref{eq5}) in dimensionless units.}, we assume equal laser detunings $\Delta$ and dipole matrix elements $\mu$ for the two transitions.

For sufficiently large detunings, $\Delta$, we can neglect the kinetic energy in Eq.(\ref{eq1a}) and adiabatically eliminate the excited-state dynamics, which gives  $\hat{\Psi}_\pm=-\frac{\mu}{2\hbar \Delta}\hat{\Psi}_g \hat{\mathcal{E}}_\pm$. Considering a quasi-2D condensate, the remaining atomic field operator can be factorized as $\hat{\Psi}_{g}(\mathbf{r},z)=\hat{\psi}(\mathbf{r})\psi_z(z)$, with $\int |\psi_z(z)|^2 {\rm d}z=1$ \cite{Castin,Petrov}. We consider a longitudinal density profile that is finite for $0<z<\ell_z$ [cf. Fig.\ref{fig1}(a)], such that $\int_0^{\ell_z} |\psi_z(z)|^2 {\rm d}z=1$. Substituting this expression and the adiabatic solution for $\hat{\Psi}_\pm$ into Eqs.(\ref{eq1b}) and (\ref{eq1c}) gives
\begin{subequations}
\begin{align}
&i\frac{\partial \hat{\psi}(\mathbf{r})}{\partial t} =-\frac{\hbar}{2m}\nabla^2_\bot \hat{\psi} + g_{2D} \hat{\psi}^\dagger\hat{\psi}\hat{\psi} +\hat{V} \hat{\psi} \label{eq2a} \\
&i \frac{\partial \hat{\mathcal{E}}_\pm (\mathbf{r},z)}{\partial z}= \mp \frac{1}{2k} (\frac{\partial^2 }{\partial z^2} + \nabla^2_\perp) \hat{\mathcal{E}}_\pm \pm \frac{k \mu^2}{2 \varepsilon_0 \hbar\Delta} |\psi_z|^2 \hat{\psi}^\dagger \hat{\psi} \hat{\mathcal{E}}_\pm  \label{eq2b}
\end{align}
\label{eq2}%
\end{subequations}
where $g_{2D}=g_{3D} \int |\psi_z(z)|^4 {\rm d}z$ is the quasi-2D interaction \cite{Castin,Petrov} and $\hat{V}=\frac{\mu^2}{4\hbar^2 \Delta}\int |\psi_z|^2( \hat{\mathcal{E}}^\dagger_+ \hat{\mathcal{E}}_+ + \hat{\mathcal{E}}^\dagger_- \hat{\mathcal{E}}_-){\rm d}z$. For a sufficiently short longitudinal extent of the BEC, we can solve the field propagation through the condensate by neglecting diffraction in Eq.(\ref{eq2b}) \cite{Camara,Robb} and obtain
\begin{equation}
\hat{\mathcal{E}}_+(\mathbf{r},z)
=e^{-i \frac{k \mu^2}{2 \varepsilon_0  \hbar\Delta} \hat{\psi}^\dagger (\mathbf{r}) \hat{\psi} (\mathbf{r}) \int^z_0 |\psi_z(z')|^2 {\rm d}z' } \hat{\mathcal{E}}_0,
\label{eq3}
\end{equation}
where $\hat{\mathcal{E}}_0$ describes the incoming light field. On the contrary, light propagation between the condensate and the mirror is entirely determined by diffraction and yields for the back-reflected amplitude at $z=\ell_z$
\begin{equation}
\hat{\mathcal{E}}_-(\mathbf{r},\ell_z)=\frac{1}{2\pi}\int \tilde{\mathcal{E}}_+(\mathbf{p}) \Phi^*(\mathbf{p}) e^{-i\mathbf{p}\cdot \mathbf{r}} {\rm d}^2 \mathbf{p},
\label{eq4}
\end{equation}
where $\tilde{\mathcal{E}}_+(\mathbf{p})$ denotes the Fourier transform of $\hat{\mathcal{E}}_+(\mathbf{r},\ell_z)$, $\Phi(\mathbf{p})=e^{2d(ik-\sqrt{\mathbf{p}^2-k^2})}$, and $d$ is the distance between the condensate and the mirror, as indicated in Fig.\ref{fig1}(a). From Eqs.(\ref{eq3}) and (\ref{eq4}) we obtain for the last term in Eq.(\ref{eq2a}) 
\begin{align}
\hat{V}(\mathbf{r})= &\frac{\Omega^2}{4 \Delta} + \frac{\Omega^2}{16\pi^2 \Delta} \left|\int e^{-i \frac{3\pi \Gamma}{2 k^2 \Delta} \hat{\psi}^\dagger(\mathbf{r}')\hat{\psi}(\mathbf{r}')}  f (\mathbf{r}-\mathbf{r}') {\rm d}^2 \mathbf{r}'\right|^2,
\label{eq4x}
\end{align}
where $\Omega=\frac{\mu \mathcal{E}_0}{\hbar}$ is the Rabi frequency of the incident field, $\Gamma=\frac{k^3 \mu^2}{3\pi \varepsilon_0 \hbar}$ denotes the spontaneous decay rate of the excited states, and $f(\mathbf{r})=\frac{1}{2\pi}\int \Phi^* (\mathbf{p}) e^{-i\mathbf{p} \cdot \mathbf{r}} {\rm d}^2\mathbf{p} $. Finally, expanding the exponent in Eq.(\ref{eq4x}), assuming $\frac{3\pi \Gamma}{2 k^2 \Delta} \langle \hat{\psi}^\dagger \hat{\psi} \rangle \ll 1$ and rotating out the constant term, we obtain a closed effective Heisenberg equation for the atoms
\begin{equation}
\begin{split}
i\frac{\partial}{\partial t}\hat{\psi}(\mathbf{r})=&-\frac{1}{2}\nabla^2_\bot \hat{\psi}(\mathbf{r}) +g\hat{\psi}^\dagger (\mathbf{r}) \hat{\psi} (\mathbf{r}) \hat{\psi}(\mathbf{r}) \\&+2\alpha\beta \int \hat{\psi}^\dagger (\mathbf{r}')\hat{\psi} (\mathbf{r}') U(\mathbf{r}-\mathbf{r}'){\rm d}^2\mathbf{r}' \hat{\psi}(\mathbf{r}),
\end{split}
\label{eq5}
\end{equation}
which contains an induced effective interaction potential $U(\mathbf{r})=2{\rm Im}[f(\mathbf{r})]$. We have scaled space and time by $\sqrt{\frac{d}{k}}$ and $\frac{m d}{\hbar k}$, respectively, giving the dimensionless parameters $ \alpha =\frac{m d \Omega^2}{16\pi \hbar k \Delta}$, $\beta = \frac{3\pi \Gamma}{2 k d \Delta}$ and $g=g_{2D}\frac{m}{\hbar}$. It follows from the above considerations, that the excited state population is negligibly small as long as $\Delta\gg\Omega$, such that $\hat{\psi}$ can be used as the atomic field operator, regardless of their electronic state.

\begin{figure}[t!]
\centering
  \includegraphics[width=0.99\columnwidth]{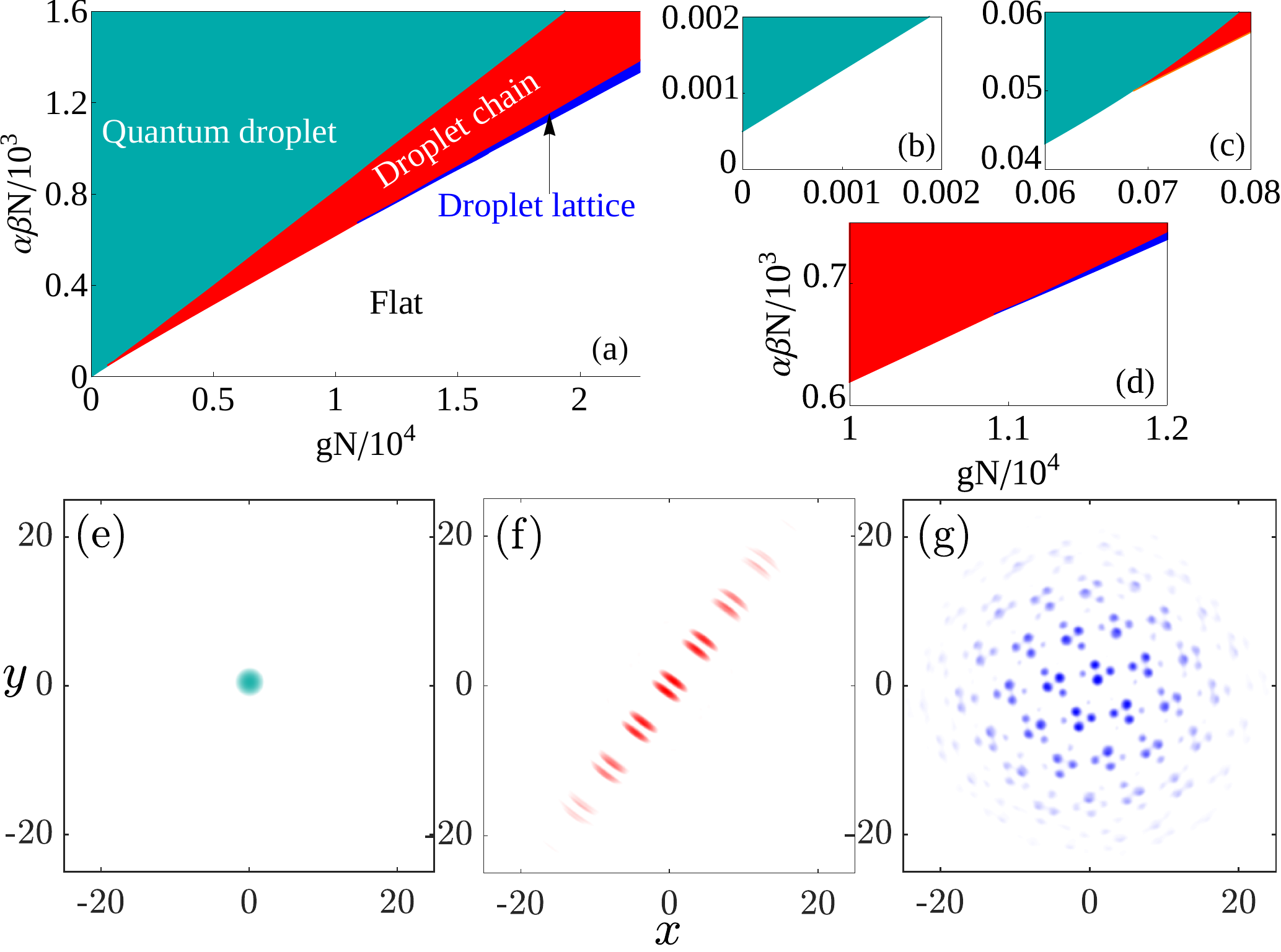}
\caption{(color online) (a-d) Ground state phase diagram of the driven BEC, determined by the competition between the zero-range and finite-range interaction with strengths $g$ and $\alpha\beta$, respectively. Panels (b-d) provide a closer look at the thresholds beyond which a new phase can be stabilized. 
Panels (e-g) illustrate the different solutions and depict the density profile of the (e) single-droplet state ($gN/10^4=0.9$), (f) droplet chain state ($gN/10^4=1.475$), and (g) droplet lattice state ($gN/10^4=1.625$) for $\alpha\beta N/10^3=1$.}
\label{fig2}
\end{figure}

The characteristic shape of the interaction is shown in Fig.\ref{fig1}(b). The potential oscillates with a growing frequency and a slowly damped amplitude as the interatomic distance increases. Simple insights can be gained for infinite mirror distances $kd\rightarrow\infty$. In this limit, the interaction potential takes on the form $U(\mathbf{r})= -\cos(r^2/4)$, which features persistent oscillations, with consecutive minima at $r_n=\sqrt{8n\pi}$ ($n=0,\ 1,\ 2, \cdots$). Even though this approximation starts to deviate significantly at large distances, it still provides a good description of the resulting BEC dynamics, as we shall see below. This is due to the finite extent of any emerging density patterns in the BEC, which tends to smear out such fine long-distance details, as illustrated in Fig.{\ref{fig1}(c). 
At short distances, the interaction is attractive and forms a potential well with a length scale $\sim \sqrt{d/k}$ that can be made significantly larger than the wavelength of the applied laser field. Interestingly, though, its Fourier transform vanishes at zero momentum such that the induced interaction potential does not promote phonon excitations in the Bogoliubov spectrum of the condensate. 

Within mean field theory, we approximate $\hat{\psi}({\bf r},t)\approx \psi({\bf r},t)=\langle\hat{\psi}({\bf r},t)\rangle$ by the condensate wave function $\psi({\bf r},t)$ that determines the particle density $|\psi({\bf r},t)|^2$ \cite{TF}. Specifically, we explore the ground states of the system by imaginary time evolution of the underlying Gross-Pitaevskii equation resulting from Eq.(\ref{eq5}) for a fixed particle number $N=\int |\psi(\mathbf{r})|^2{\rm d}^2 \mathbf{r}$. Depending on the two parameters $\alpha\beta N$ and $gN$, we find different ground states [Fig.\ref{fig2}(a)], from an infinitely extended flat BEC for strong zero-range interactions to a single localized quantum droplet [cf. Fig.\ref{fig2}(e)] for dominating non-local interactions, $U(\mathbf{r})$. 
In between these limiting cases, there exist two-dimensional regular structures as exemplarily shown in Fig.\ref{fig2}(g), and one-dimensional droplet chains as depicted in Fig.\ref{fig2}(f).

\begin{figure}[b!]
\centering
  \includegraphics[width=0.99\columnwidth]{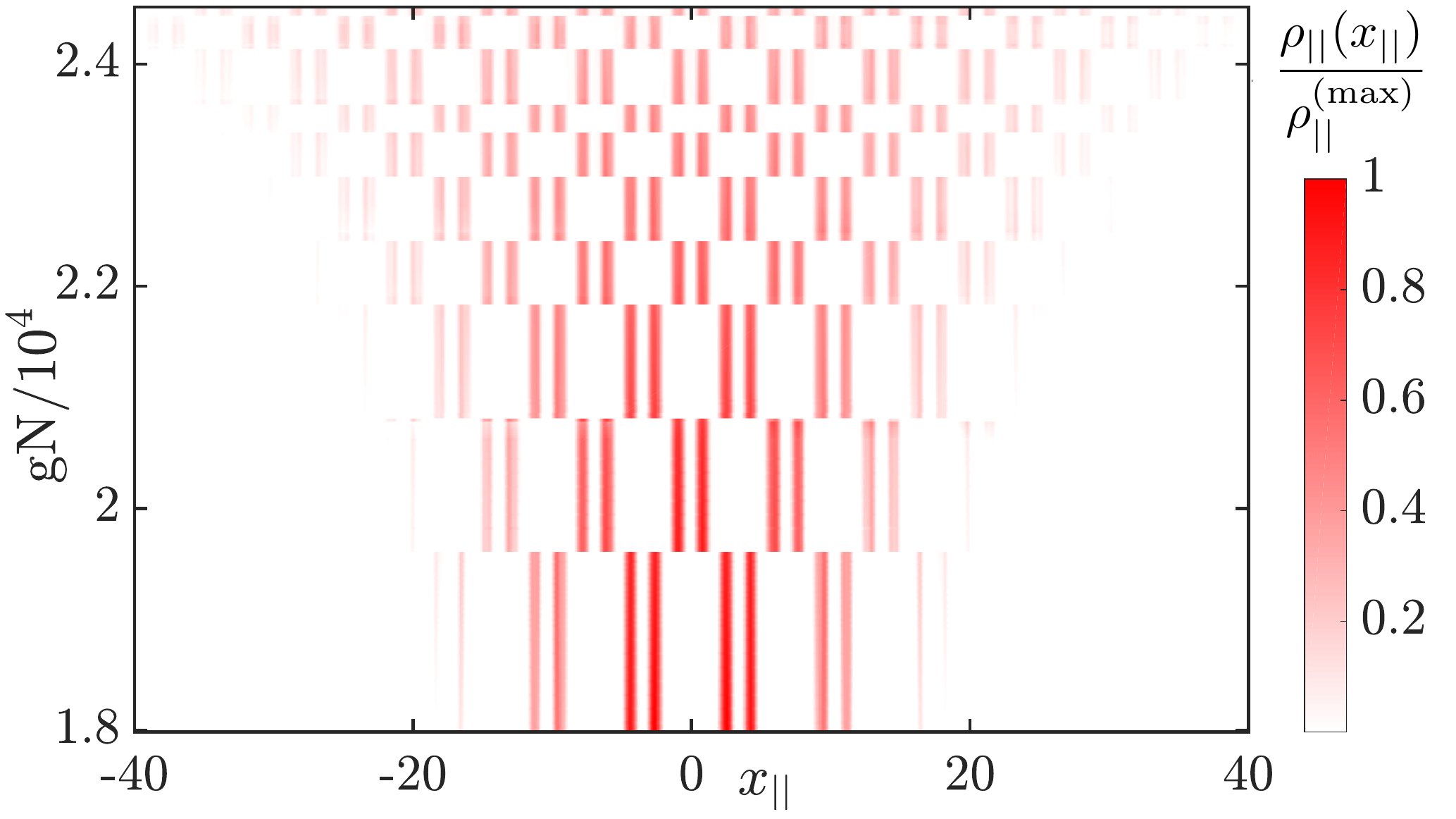}
\caption{(color online) Longitudinal density profile $\rho_{||}(x_{||})$ in the droplet-chain phase with increasing strength $g$ of the zero-range interaction for $\alpha\beta N/10^3=1.5$. All profiles are normalized by the maximum density $\rho_{||}^{(\rm max)}$ at the smallest interaction strength $gN/10^4=1.8$.}
\label{fig3}
\end{figure}

The formed quantum droplet is bound by the short-distance well of the induced interaction potential [cf. Fig.\ref{fig1}(b)] and therefore has a typical size of $\sim \sqrt{d/k}$, i.e., of order $\sim 1$ in the used scaled units. As the strength $g$ of the zero-range interaction increases, the size of the droplet increases beyond the first minimum of the finite-range interaction potential, which eventually leads to the formation of richer structures. Starting from the single-droplet solution, we see that the one-dimensional droplet chains appear in a substantial range of parameters in phase diagram [Fig.\ref{fig2}(a)], and are energetically favoured by the long-range interaction as compared to extended 2D droplet arrays. The particular $\sqrt{n}$-scaling of the location of the $n$th potential minimum leads to a local double-peak structure, resembling a dimer chain, which matches the position of the first few potential minima as illustrated in Fig.\ref{fig1}(d). The found emergence of a one-dimensional structure out of a 2D geometry is indeed remarkable considering the isotropic nature of the interaction potential, and can be traced back to its long-ranged oscillatory form. Upon further increase of $g$, local repulsion starts to dominate and eventually leads to a completely delocalized structureless state with constant density. In between these two phases, we find two-dimensional structures with broken translational symmetry in the form of two-dimensional droplet lattices. Similar to the droplet chains, each site features a finer structure that emerges from the particular $\sqrt{n}$-scaling of the potential minima as discussed above. 

Returning to the one-dimensional structure, the size of the emerging droplet chain is controlled by both interaction parameters. Upon increasing $g N$, the droplet chain grows with a stepwise rising number of sites as shown in Fig.\ref{fig3}. Here, $\rho_\parallel(x_\parallel)$ is the longitudinal density, obtained by integrating the 2D density $|\psi({\bf r})|^2$ over the coordinate transverse to the chain axis. As the strength $g N$ of the local repulsion is increased, the site number grows, alternating between even and odd chain sizes, accompanied by a gradual decrease of the overall density. Beyond a critical length, the chain eventually dissolves into the delocalized state or is replaced by the 2D droplet lattice state discussed above. By simultaneously increasing the strength $\alpha \beta N$ of the light-induced interaction close to this phase boundary, the chain keeps extending continually and approaches the thermodynamic limit as $N\rightarrow\infty$. Note, however, that a sole increase of the atom number $N$ for fixed values of $\alpha\beta$ and $g$ drives the system into the Thomas-Fermi limit in which kinetic energy becomes irrelevant and the formed structure is entirely determined by the competition of the two interaction terms. In this case, the emerging density profile becomes independent of $N$ and a further increase of the particle number simply leads to a linear growth of the overall density.

\begin{figure}[t!]
\centering
  \includegraphics[width=0.99\columnwidth]{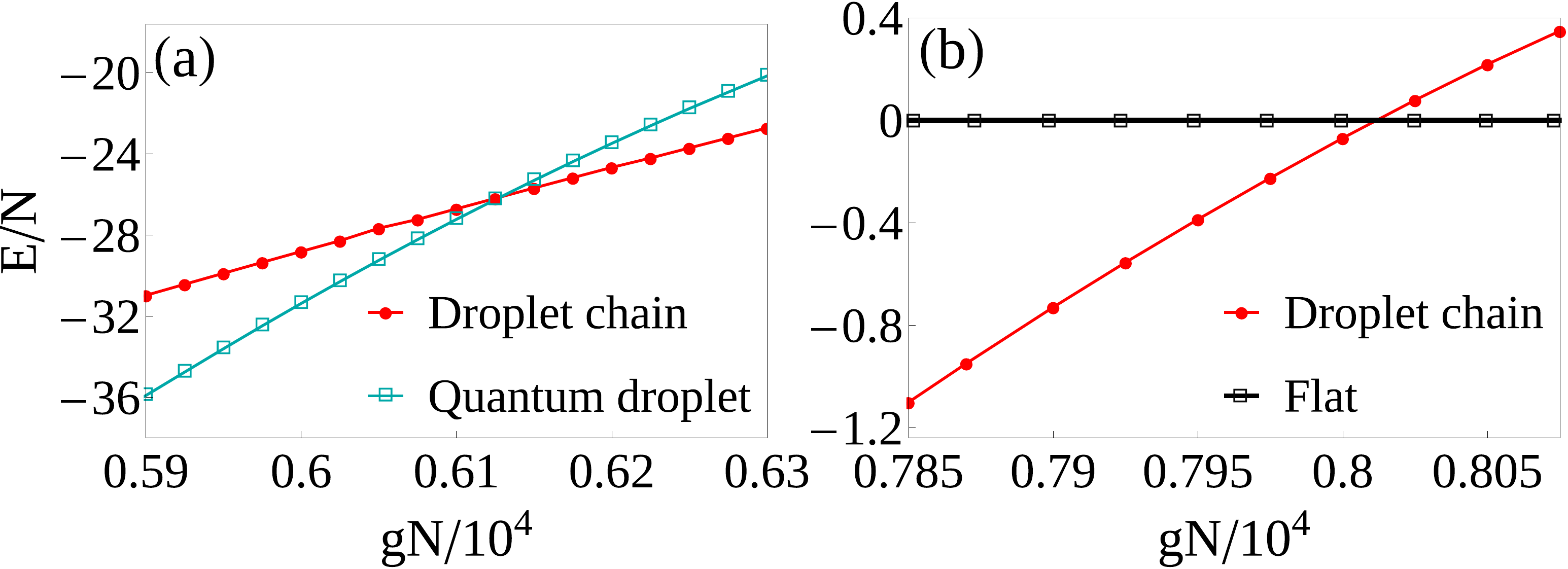}
\caption{(color online) Ground state energy of different indicated solutions as a function of the strength of the zero-range interaction for $\alpha\beta N/10^3=0.5$.}
\label{fig4}
\end{figure}

Transitions between the different states are, however, driven by the kinetic energy. As shown in Fig.\ref{fig4}, comparing the total energies of different states as a function of $g N$ while keeping $\alpha\beta N$ fixed, we know that the found transitions are of first order. The appearance of all nontrivial phases involves a threshold strength of the nonlocal light-induced interaction. While the formation of the self-bound quantum droplet only requires a relatively small value of $\alpha\beta N\approx0.5$ for vanishing local repulsion $g=0$ [Fig.\ref{fig2}(b)], the emergence of the droplet chain and droplet lattice states requires significantly larger values of $\alpha\beta N\approx 50$ and $\alpha\beta N\approx 675$, as shown in Figs.\ref{fig2}(c) and (d), respectively.

\begin{figure}[t!]
\centering
  \includegraphics[width=0.99\columnwidth]{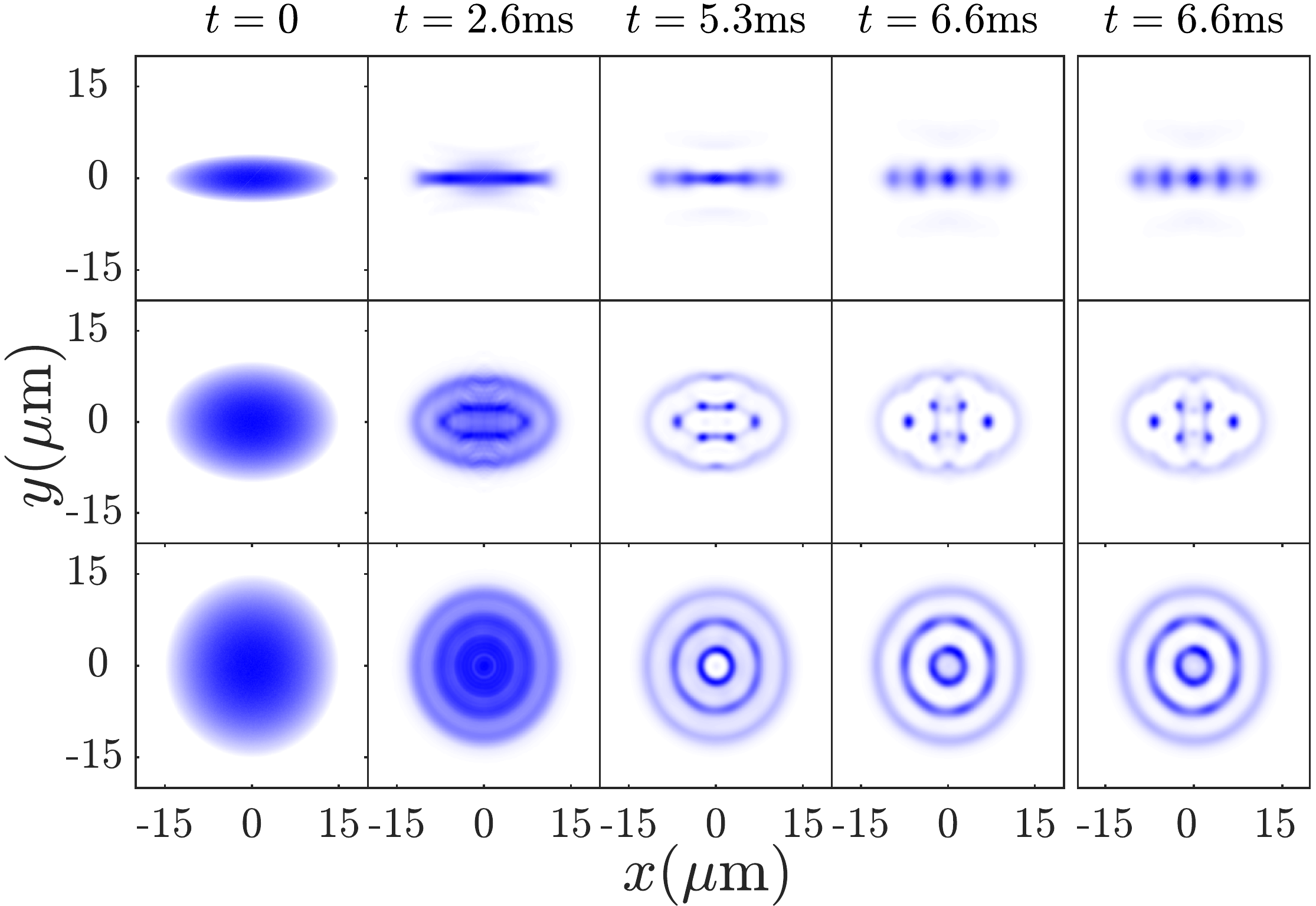}
\caption{(color online) Time evolution of a laser-driven rubidium BEC (from left to right) for different initial density profiles depicted in the leftmost column.  The simulations account for scattering-induced atom loss and are performed for $\Omega/\Delta=2.5\times 10^{-3}$, $d=10\lambda$, $\lambda=780{\rm nm}$, $a=0.62{\rm nm}$ and an initial peak density of $\rho_0=2.8\times 10^{14} {\rm cm}^{-3}$. The rightmost column shows the final density for the infinite-range interaction in the $d/\lambda\rightarrow\infty$ limit, and is virtually identical to the result for $d=10\lambda$.
}
\label{fig5}
\end{figure}

In order to explore observable signatures of the described states in potential experiments, we have studied the real dynamics of the system under relevant experimental conditions. The main challenge stems from spontaneous photon emission on the two optically driven transitions, which leads to condensate loss and thereby limits the observation time. In the considered limit of a large laser detuning, the dissipation rate can be determined from $\Gamma_{\rm sc}=\frac{\Gamma \Omega^2}{2\Delta^2}$, i.e., the decay rate $\Gamma$ multiplied by the small excited-state population $\frac{\Omega^2}{2\Delta^2}$. 
Within our scaled units, the dimensionless condensate loss rate is thus given by $\tilde{\Gamma}_{\rm sc} =\Gamma_{\rm sc}\frac{md}{\hbar k}=16kd\alpha\beta/3$. The corresponding timescale for condensate loss should be significantly longer than the typical dynamical time for structure formation. Since the latter scales with the interaction strength $\alpha\beta N$ a high atomic density provides favourable conditions. Moreover, the actual parameters are chosen such that $|\Delta| \gg \Gamma,\Omega$; $\ell_z<d$ and $\beta |\psi(\mathbf{r})|^2 =\frac{3 \pi^{3/2} \Gamma}{2 k^2 \Delta} \rho_0 \ell_z\ll 1$ to ensure the validity of the effective theory Eq.(\ref{eq5}). In the last inequality, $\rho_0$ denotes the atomic peak density. 

These constraints can be satisfied without fundamental difficulty as demonstrated in Fig.\ref{fig5}, where we show the dynamics of a quasi-2D rubidium BEC in the presence of atom losses under realistic experimental conditions \cite{Bloch1,Petrov}. Starting from an initial Thomas-Fermi profile \cite{TF} $|\psi(t=0)|^2=\rho_0 (1-\frac{x^2}{\sigma^2_x}-\frac{y^2}{\sigma^2_y}) e^{-z^2/\ell^2_z}$ and $\ell_z=1\mu$m the system indeed undergoes a rapid dynamical formation of structures on a timescale of a few milliseconds before atom loss starts to significantly deplete the condensate and decrease its density. Depending on the initial aspect ratio, $\sigma_x/\sigma_y$, the BEC relaxes towards different states that reflect the ground state physics discussed above. 

In conclusion, we have shown that coupling light to a BEC within a single-mirror feedback configuration leads to interesting effective interactions between the atoms. Its long range and oscillatory behaviour stands out from other types of interactions that are currently considered in ultracold atomic quantum gases \cite{saffman_rev,pfau1}. This peculiar shape was found to cause surprising behavior such as a transition to one-dimensional chains of quantum droplets from a two-dimensional condensate. Recently, there has been significant work on the formation of quantum droplets and their regular structures \cite{pfau1,pfau2}. The present 2D system provides the first instance that promotes such regular structures in a self-stabilized fashion without external confinement. 
Our simulations show that this behavior should be within experimental reach which, in light of the comparable technological simplicity of the considered setup, suggest an intriguing alternative route to long-range interactions in atomic quantum gases. Given the peculiar nature of the found effective interactions, the consequences of strong laser driving beyond the small-phase regime, effects of strong interactions and quantum fluctuations as well as the possibility of supersolidity in the present system all present exciting future questions to be explored.

We are grateful to Francesco Piazza, Thorsten Ackemann and Nils Byg J\o{}rgensen for fruitful discussions and helpful comments. 
This work was funded by the Danish National Research Foundation through a Niels Bohr Professorship.


\begin{thebibliography}{99}

\bibitem{Bloch1}
I. Bloch, J. Dalibard, and W. Zwerger, Rev. Mod. Phys. \textbf{80}, 885 (2008).

\bibitem{Gross}
C. Gross, and I. Bloch, Science \textbf{357}, 995 (2017).

\bibitem{Polkovnikov}
A. Polkovnikov, K. Sengupta, A. Silva, and M. Vengalattore, Rev. Mod. Phys. \textbf{83}, 863 (2011).

\bibitem{Nandkishore}
R. Nandkishore, and D. A. Huse, Annu. Rev. Condens. Matter Phys. \textbf{6}, 15 (2015).

\bibitem{Langen}
T. Langen, R. Geiger, and J. Schmiedmayer, Annu. Rev. Condens. Matter Phys. \textbf{6}, 201 (2015).

\bibitem{Chin}
C. Chin, R. Grimm, P. Julienne, and E. Tiesinga, Rev. Mod. Phys. \textbf{82}, 1225 (2010).

\bibitem{roos}
R. Blatt, and C. F. Roos, Nat. Phys. \textbf{8}, 277 (2012).

\bibitem{Carr}
L. D. Carr, D. DeMille, R. V. Krems, and J. Ye, New J. Phys. \textbf{11}, 055049 (2009).

\bibitem{Yan}
B. Yan, S. A. Moses, B. Gadway, J. P. Covey, K. R. A. Hazzard, A. M. Rey, D. S. Jin, and J. Ye, Nature (London), \textbf{501}, 521 (2013).

\bibitem{saffman_rev}
M. Saffman, T. G. Walker, and K. M\o{}lmer, Rev. Mod. Phys. {\bf 82}, 2313 (2010).

\bibitem{santos2003}
L. Santos, G. V. Shlyapnikov, and M. Lewenstein, Phys. Rev. Lett. {\bf 90}, 250403 (2003).

\bibitem{schauss2012}
P. Schau\ss, M. Cheneau, M. Endres, T. Fukuhara, S. Hild, A. Omran, T. Pohl, C. Gross, S. Kuhr, and I. Bloch, Nature (London) {\bf 491}, 87 (2012).

\bibitem{schauss2015}
P. Schau\ss, J. Zeiher, T. Fukuhara, S. Hild, M. Cheneau, T. Macr\`{i}, T. Pohl,  I. Bloch, and C. Gross, Science {\bf 347}, 1455 (2015).

\bibitem{Zeiher}
J. Zeiher, R. van Bijnen, P. Schau\ss, S. Hild, J. Choi, T. Pohl, I. Bloch, and C. Gross, Nat. Phys. \textbf{12}, 1095 (2016).

\bibitem{Labuhn}
H. Labuhn, D. Barredo, S. Ravets, S. de L$\rm \acute{e}$s$\rm \acute{e}$leuc, T. Macr\`{i}, T. Lahaye, and A. Browaeys, Nature (London) \textbf{534}, 667 (2016).

\bibitem{zeiher2017}
J. Zeiher, J. Y. Choi, A. Rubio-Abadal, T. Pohl, R. van Bijnen, I. Bloch, and C. Gross, Phys. Rev. X {\bf 7}, 041063 (2017).

\bibitem{Bernien}
H. Bernien, S. Schwartz, A. Keesling, H. Levine, A. Omran, H. Pichler, S. Choi, A. S. Zibrov, M. Endres, M. Greiner, V. Vuleti\'{c}, and M. D. Lukin, Nature (London) \textbf{551}, 579 (2017).

\bibitem{pfau1}
H. Kadau, M. Schmitt, M. Wenzel, C. Wink, T. Maier, I. Ferrier-Barbut, and T. Pfau, Nature (London) {\bf 530}, 194 (2016).

\bibitem{pfau2}
I. Ferrier-Barbut, H. Kadau, M. Schmitt, M. Wenzel, and T. Pfau, Phys. Rev. Lett. {\bf 116}, 215301 (2016).

\bibitem{ferlaino2016}
L. Chomaz, S. Baier, D. Petter, M. J. Mark, F. W\"achtler, L. Santos, and F. Ferlaino, Phys. Rev. X {\bf  6}, 041039 (2016).

\bibitem{Devreese}
J. T. Devreese, and A. S. Alexandrov, Rep. Prog. Phys. \textbf{72}, 066501 (2009).

\bibitem{Camacho1}
A. Camacho-Guardian, and G. M. Bruun, arXiv:1712.06931 (2017).

\bibitem{Camacho2}
A. Camacho-Guardian, L. A. Pe\~{n}a Ardila, T. Pohl, and G. M. Bruun, arXiv:1804.00402v1 (2018).

\bibitem{Petersen}
J. Petersen, J. Volz, and A. Rauschenbeutel, Science {\bf 346}, 67 (2014).

\bibitem{Douglas}
J. S. Douglas, H. Habibian, C.-L. Hung, A. V. Gorshkov, H. J. Kimble, and D. E. Chang, Nat. Photon. {\bf 9}, 326 (2015).

\bibitem{Baumann}
K. Baumann, C. Guerlin, F. Brennecke, and T. Esslinger, Nature (London) \textbf{464}, 1301 (2010).

\bibitem{Mottl}
R. Mottl, F. Brennecke, K. Baumann, R. Landig, T. Donner, and T. Esslinger, Science \textbf{336}, 1570 (2012).

\bibitem{Habibian}
H. Habibian, A. Winter, S. Paganelli, H. Rieger, and G. Morigi, Phys. Rev. Lett. \textbf{110}, 075304 (2013).

\bibitem{Schmidt}
D. Schmidt, H. Tomczyk, S. Slama, and C. Zimmermann, Phys. Rev. Lett. \textbf{112}, 115302 (2014).

\bibitem{Jager}
S. B. J\"{a}ger, S. Sch\"{u}tz, and G. Morigi, Phys. Rev. A \textbf{94}, 023807 (2016).

\bibitem{Landig}
R. Landig, L. Hruby, N. Dogra, M. Landini, R. Mottl, T. Donner, and T. Esslinger, Nature (London) \textbf{532}, 476 (2016).

\bibitem{Flottat}
T. Flottat, L. de Forges de Parny, F. H\'{e}bert, V. G. Rousseau, and G. G. Batrouni, Phys. Rev. B \textbf{95}, 144501 (2017).

\bibitem{Leonard}
J. L\'{e}onard, A. Morales, P. Zupancic, T. Esslinger, and T. Donner, Nature (London) \textbf{543}, 87 (2017).

\bibitem{Vaidya}
V. D. Vaidya, Y. Guo, R. M. Kroeze, K. E. Ballantine, A. J. Koll\'{a}r, J. Keeling, and B. L. Lev, Phys. Rev. X \textbf{8}, 011002 (2018).

\bibitem{Piazza}
F. Piazza, and H. Ritsch, Phys. Rev. Lett. \textbf{115}, 163601 (2015).

\bibitem{Mivehvar}
F. Mivehvar, S. Ostermann, F. Piazza, and H. Ritsch, Phys. Rev. Lett. {\bf 120}, 123601 (2018).

\bibitem{Ostermann}
S. Ostermann, F. Piazza, and H. Ritsch, Phys. Rev. X \textbf{6}, 021026 (2016).

\bibitem{Schmittberger}
B. L. Schmittberger, and D. J. Gauthier, New J. Phys. \textbf{18}, 103021 (2016).

\bibitem{Labeyire}
G. Labeyrie, E. Tesio, P. M. Gomes, G.-L. Oppo, W. J. Firth, G. R. M. Robb, A. S. Arnold, R. Kaiser, and T. Ackemann, Nat. Photon. \textbf{8}, 321 (2014).

\bibitem{Ackemann}
T. Ackemann, and W. Lange, Phys. Rev. A \textbf{50}, 4468(R) (1994).

\bibitem{Tesio}
E. Tesio, G. R. M. Robb, G.-L. Oppo, P. M. Gomes, T. Ackemann, G. Labeyire, R. Kaiser, and W. J. Firth, Phil. Trans. R. Soc. A \textbf{372}, 20140002 (2014).

\bibitem{Tesio1}
E. Tesio, G. R. M. Robb, T. Ackemann, W. J. Firth, and G.-L. Oppo, Phys. Rev. Lett. \textbf{112}, 043901 (2014).

\bibitem{Firth3}
W. J. Firth, I. Kre\v{s}i\'{c}, G. Labeyrie, A. Camara, and T. Ackemann, Phys. Rev. A \textbf{96}, 053806 (2017).

\bibitem{Robb}
G. R. M. Robb, E. Tesio, G.-L. Oppo, W. J. Firth, T. Ackemann, and R. Bonifacio, Phys. Rev. Lett. \textbf{114}, 173903 (2015).

\bibitem{Camara}
A. Camara, R. Kaiser, G. Labeyrie, W. J. Firth, G.-L. Oppo, G. R. M. Robb, A. S. Arnold, and T. Ackemann, Phys. Rev. A \textbf{92}, 013820 (2015).

\bibitem{Moriya}
P. H. Moriya, R. F. Shiozaki, R. Celistrino Teixeira, C. E. M\'{a}ximo, N. Piovella, R. Bachelard, R. Kaiser, and Ph. W. Courteille, Phys. Rev. A \textbf{94}, 053806 (2016).

\bibitem{Cinti1}
F. Cinti, P. Jain, M. Boninsegni, A. Micheli, P. Zoller, and G. Pupillo, Phys. Rev. Lett. {\bf 105}, 135301 (2010).

\bibitem{Hsueh}
C.-H. Hsueh, T.-C. Lin, T.-L. Horng, and W. C. Wu, Phys. Rev. A \textbf{86}, 013619 (2012).

\bibitem{Cinti2}
F. Cinti, T. Macr\`{i}, W. Lechner, G. Pupillo, and T. Pohl, Nat. Comm. \textbf{5}, 3235 (2014).

\bibitem{Viteau}
M. Viteau, M. G. Bason, J. Radogostowicz, N. Malossi, D. Ciampini, O. Morsch, and E. Arimondo, Phys. Rev. Lett. \textbf{107}, 060402 (2011).

\bibitem{QuantumOpt}
M. O. Scully, and M. S. Zubairy, \emph{Quantum Optics} (Cambridge University Press, 2001).

\bibitem{Castin}
Y. Castin, and R. Dum, Eur. Phys. J. D \textbf{7}, 399 (1999).

\bibitem{Petrov}
D. S. Petrov, M. Holzmann, and G. V. Shlyapnikov, Phys. Rev. Lett. \textbf{84}, 2551 (2000).

\bibitem{TF}
C. J. Pethick, and H. Smith, \emph{Bose-Einstein Condensation in Dilute Gases} (Cambridge University Press, 2008).

\end{thebibliography}
\end{document}